 \documentclass[final,3p,times]{elsarticle}
\usepackage{amssymb}
\usepackage{amsthm}

\usepackage{lineno}

\begin{document}

\begin{frontmatter}



\title{The $R_{uds}$ value in the vicinity of $\psi(3770)$ state}
\author{
Rong Wang$^{a,b,c}$,
Xu Cao$^{a,d}${\footnote{Corresponding author: caoxu@impcas.ac.cn}},
Xurong Chen$^{a}$
}
\address{
$^a$ Institute of Modern Physics, Chinese Academy of Sciences, Lanzhou 730000, China \\
$^b$ Lanzhou University, Lanzhou 730000, China \\
$^c$ University of Chinese Academy of Sciences, Beijing 100049, China \\
$^d$State Key Laboratory of Theoretical Physics, Institute of Theoretical Physics, Chinese
Academy of Sciences, Beijing 100190, China
}

\begin{abstract}
The anomalous line shape of the $\psi(3770)$ state has resulted in some difficulty
in the determination of the $R$ value for the continuum light hadron production
in the resonance energy range. We parameterize the asymmetric line shape
using a Fano-type formula and extract the $R_{uds}$ value to be $2.156\pm 0.022$
from the data of BESIII Collaboration in the energy region between 3.650 and 3.872 GeV.
The small discrepancy between experiment and theory is removed. The cross sections of
the $e^+e^- \rightarrow hadrons$ are given by subtracting the part of continuum light
hadron production and are compared to the data of the $e^+e^- \rightarrow D\bar{D}$ reaction.
\end{abstract}


\end{frontmatter}


\section{Introduction} \label{sec:intro}

The cross section of the $e^+e^- \rightarrow hadrons$ in terms of the center-of-mass (c.m.)
energy is one of the most fundamental observables in Quantum Chromodynamics (QCD).
The final hadrons are produced via a pair of quark-antiquark proceeded
from a virtual photon by initial-state electron-positron annihilation.
Instead of the cross section for inclusive hadron production,
the hadronic R-ratio $R(s)$ is often used owning to its simplicity
both on the experimental and the theoretical side,
\begin{equation}
R(s)=\frac{\sigma(e^+e^-\rightarrow hadrons)}{\sigma(e^+e^-\rightarrow \mu^+\mu^-)} \quad,
\end{equation}
where $\sigma(e^+e^-\rightarrow \mu^+\mu^-)=4\pi\alpha^2/3s$ is the photon-mediated
lowest order muon pair production cross section with $s$ being the square of c.m. energy
and $\alpha$ the electromagnetic coupling constant. If no resonances are present,
the $R(s)$ values solely from the continuum hadrons are well known to be
$3\sum_{f}Q^2_f$ in the lowest order approximation, with $f$ being quark flavors
and $Q_f$ the corresponding quark charge. The higher order corrections from
the finite quark masses and the gluonic emission could be calculated by perturbative QCD
(pQCD)~\cite{A.D.Martin,rhad,P.A.Baikov}. So the measurement of $R(s)$ is important
for testing the validity of both pQCD calculation
and hadronic vacuum polarization correction.

The $R(s)$ for the continuum light hadron (containing $u$, $d$ and $s$ quarks) production,
denoted as $R_{uds}$ in this letter, is usually used to test
the validity of the pQCD calculation in relatively low energy region.
Precise measurements of the $R_{uds}$ near the $D\bar{D}$ threshold are reported by
BES Collaboration \cite{BES-prl97-262001,BES-prl97-121801,BES-plb652,BES-plb641}.
The $R_{uds}$ value below the $D\bar{D}$ threshold is not affected by
the first $D\bar{D}$ open charm resonance $\psi(3770)$, therefore, determination of
$R_{uds}$ is very simple for the case. The $R_{uds}$ in the energy region from 3.650 to 3.732 GeV is determined
to be $R_{uds}=2.141 \pm 0.025 \pm 0.085$ \cite{BES-prl97-262001},
which is in good agreement with $R^{pQCD}_{uds}=2.15$ predicted
by pQCD \cite{A.D.Martin,rhad,P.A.Baikov}.
However, the $R_{uds}$ value in the open charm threshold region is overlapped
with many resonances. The obtained value varies widely among different fits,
e.g. in or nearby the $\psi(3770)$ resonance. It is extracted to be
$R_{uds}=2.262 \pm 0.054 \pm 0.109$ in the energy region from 3.660 to 3.872 GeV
\cite{BES-prl97-121801} and to be $R_{uds}=2.121 \pm 0.023 \pm 0.084$ in the energy region
from 3.650 to 3.872 GeV~\cite{BES-plb652}.
There are obvious differences among the obtained $R_{uds}$ values
and also the pQCD calculation. As a matter of fact, these extracted $R_{uds}$
values depend on the treatment of resonances in the $\psi(3770)$ region.
Therefore, a more reliable method is required to reasonably
extract the $R_{uds}$ value from the experimental data.

In order to accurately extract the $R_{uds}$ in the region of $\psi(3770)$ resonance,
the anomalous line shape of the $\psi(3770)$ state should be treated carefully.
It has been found at the very beginning that the total cross sections of
$e^+e^- \rightarrow hadrons$ in the energy range between 3.700 and 3.872 GeV
could not be described well with only one Breit-Wigner (BW) resonance
even using the energy-dependent width of the $\psi(3770)$
\cite{BES-prl97-262001,BES-prl97-121801,BES-plb652,BES-plb641}.
This is confirmed by the inclusive measurements of the
$e^+e^- \rightarrow D\bar{D}$, $D^+D^-$, $D^0\bar{D}^0$ reactions
in the similar c.m. energy region \cite{BES-prl101,BES-plb668}.
In the analysis of the BES Collaboration, a modified form of the energy-dependent width
is usually used in their fits to data,
\begin{equation}
\Gamma_{D\bar{D}}(s) \propto \frac{p^3_{0,\pm}(s)}{1 + r^2 p^2_{0,\pm}(s)} \quad ,\label{eq:BESwidth}
\end{equation}
with $p_{0,\pm}=\sqrt{s/4-m^2_{D^{0,\pm}}}$ being the final $D-$meson momentum in the c.m. system
and $r \sim 1.0$ fm the interaction radius of the $c\bar{c}$.
The BW resonance with the width in Eq.~(\ref{eq:BESwidth}) could give an asymmetric line shape
of the $\psi(3770)$ state, but does not describe well the dip around 3.82 GeV.

The deviation from the BW resonance of the $\psi(3770)$ state has inspired a lot of
interesting theoretical efforts \cite{YRLiu,HBLi,YJZhang,Achasov,Achasov2013,GYChen,Limphirat,CaoFano2014,Caopsi2014}.
Experimental measurements from both BES and CLEO Collaborations show that the $\psi(3770)$
resonance mainly decays to $D\bar{D}$ channel though the specific decay ratio is still under discussion \cite{CLEO-prl96-032003,CLEO-prl96-092002,BES-prd76,BES-plb659,CLEO-prd76,CLEO-prd89}.
However, the rescattering of final $D\bar{D}$ is found to be not enough to
account for its line shape deviation \cite{YRLiu}.
Now it is explained to be the consequence of the interference between the
$\psi(3770)$ resonance and the continuum background from the $\psi(2S)$ contribution
\cite{HBLi,YJZhang,Achasov,Achasov2013,GYChen,Limphirat}.
Its implication to the nature of $\psi(3770)$ state is also investigated
in the Fano mechanism \cite{CaoFano2014,Caopsi2014}.
In the Fano theory, the asymmetric line shape of states is produced
by the interference of continuum and resonance, which is giving rise to
a general physical phenomenon in many quantum system, e.g. the
nuclear, atomic, condensed matter physics and molecular spectroscopy.
Though the underlying physics of the $\psi(3770)$ state is still waiting for
further exploration \cite{CaoFano2014,Caopsi2014}, the Fano-type formula provides
an appropriate and simple parameterized expression for describing the
anomalous line shape of the total cross sections of the $e^+e^- \rightarrow hadrons$ and $D\bar{D}$.
In this letter, we will use this formula to extract the $R_{uds}$ value
from experimental data reported by BES Collaboration in the energy region between
3.650 and 3.872 GeV\cite{BES-prl97-262001}.

\section{Method and Result}

\begin{figure}[htp]
\centering
\includegraphics[width=0.5\textwidth]{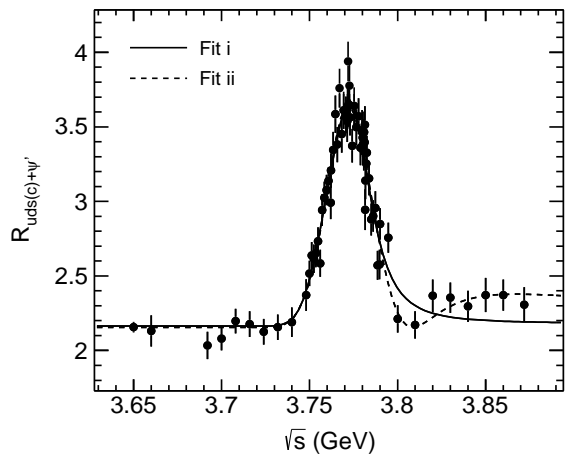}
\caption{
$R_{uds(c)+\psi^{'}}(s)$ at different c.m. energies. The curves are the fits to the
data with (solid line) and without (dashed line) $\Gamma_{\psi^{'}e^{+}e^{-}}$ fixed
to the average value in PDG. The data are measured by BESIII
Collaboration \cite{BES-prl97-262001}.
}
\label{fig1}
\end{figure}

The theoretical $R_{uds(c)+\psi^{'}}(s)$ contains the contributions from
continuum light hadron production $R_{uds}(s)$, the continuum $c\bar{c}$ production
$R_{(c)}(s)$, and the bare $\psi^{'}$ resonance production (here and below,
the $\psi(3770)$ is denoted as $\psi^{'}$ for short), which is written as
\begin{equation}
R^{th}_{uds(c)+\psi^{'}}(s)=R_{uds}(s)+R_{(c)}(s)+
\frac{\sigma(e^+e^-\rightarrow \psi^{'} \rightarrow hadrons)}{\sigma(e^+e^-\rightarrow \mu^+\mu^-)} \label{eq:Rseperate1} \quad,
\end{equation}
with $R_{(c)}(s) = f_{(c)} p^3_{0,\pm}/E^3_{0,\pm}$ in BESIII's fit \cite{BES-prl97-262001}.
The $\sigma(e^+e^-\rightarrow \psi^{'} \rightarrow hadrons)$ are the hadrons production
cross section through the bare $\psi^{'}$ resonance in $e^+e^-$ annihilation,
and it could be written in terms of the form factor $F_{\psi^{'}}(s)$ in the following way:
\begin{equation}
\sigma(e^+e^-\rightarrow \psi^{'} \rightarrow hadrons)=
\frac{8\pi\alpha^{2}}{3s^{5/2}}[p_0^3(s)+p_{\pm}^3(s)]|F_{\psi^{'}}(s)|^2 \label{eq:cshadrons1} \quad,
\end{equation}
where besides the factor from phase space, the bare $\psi^{'}$ form factor $F_{\psi^{'}}(s)$
would be taken as the BW form:
\begin{equation}
F_{\psi^{'}}(s)=\frac{g_{\psi^{'}D\bar{D}}g_{\psi^{'}\gamma}}
{s-m^2_{\psi^{'}}+im_{\psi^{'}}\Gamma_{\psi^{'}}(s)} \quad, \label{eq:bareFF}
\end{equation}
where $g_{\psi^{'}D\bar{D}}$ and $g_{\psi^{'}\gamma}$ are the coupling constants of the $\psi^{'}$
to the $D\bar{D}$ and photon, respectively. Experiments indicate that the dominated decay channel
of $\psi^{'}$ resonance is the $D\bar{D}$. Hence, we may use the energy dependent width
\begin{equation}
\Gamma_{\psi^{'}}(s)= \Gamma_{D\bar{D}} + \Gamma_{nonD\bar{D}} = g^2_{\psi^{'}D\bar{D}}\frac{p_0^3(s)+p_{\pm}^3(s)}{6\pi s} + \Gamma_{nonD\bar{D}}\quad, \label{eq:EDwidth}
\end{equation}
or an improved parameterization of $\Gamma_{D\bar{D}}$ in Eq.~(\ref{eq:BESwidth}).
However, as we addressed in Sec.~\ref{sec:intro}, Eq.~(\ref{eq:bareFF}) is enough
to describe the asymmetric line shape of the $\psi(3770)$ state,
but does not describe well the dip around 3.82 GeV. The main weakness of above treatment
is the totally separation of the continuum and resonant $D\bar{D}$ production
in Eq. (\ref{eq:Rseperate1}), but in fact, they are convoluted to each other.
This is justified by the BESIII's fit results with the lower limit of $f_{(c)} \sim 0$
within the uncertainties. Keeping this in mind, we correct the Eq. (\ref{eq:Rseperate1}) as,
\begin{equation}
R^{th}_{uds(c)+\psi^{'}}(s)=R_{uds}(s)+
\frac{\sigma(e^+e^-\rightarrow (c\bar{c})+\psi^{'} \rightarrow hadrons)}{\sigma(e^+e^-\rightarrow \mu^+\mu^-)} \label{eq:Rseperate2} \quad,
\end{equation}
The $\sigma(e^+e^-\rightarrow (c\bar{c})+\psi^{'} \rightarrow hadrons)$ are
the hadrons production cross section through the continuum $c\bar{c}$ and the $\psi^{'}$ resonance
in $e^+e^-$ annihilation, which should be alike to Eq. (\ref{eq:cshadrons1}) :
\begin{equation}
\sigma(e^+e^-\rightarrow (c\bar{c})+\psi^{'} \rightarrow hadrons)=
\frac{8\pi\alpha^{2}}{3s^{5/2}}[p_0^3(s)+p_{\pm}^3(s)] |F_{(c)+\psi^{'}}(s)|^2 \label{eq:cshadrons2} \quad,
\end{equation}
Instead of Eq.~(\ref{eq:bareFF}), the Fano-type form factor including
the interference between resonance and continuum background could be written
as \cite{CaoFano2014,Caopsi2014,Fano1961}
\begin{equation}
|F_{(c)+\psi^{'}}(s)|^2=|g_{\psi^{'}D\bar{D}}g_{\psi^{'}\gamma}F_{(c)}|^2
\frac{|q+\varepsilon|^2}{1+\varepsilon^2} \quad, \label{eq:FanoFF}
\end{equation}
with $\varepsilon=(-s+m^2_{\psi^{'}})/(m_{\psi^{'}}\Gamma_{\psi^{'}})$.
In the present context, the Fano line shape parameter $q$ characterizes
the relative transition strength into the $\psi^{'}$ state versus
the $D\bar{D}$ continuum and can be related to the electromagnetic transition
probability of the $\psi^{'}$ state. It is an energy dependent variable
in the original formula but regraded as a constant in the limited energy range
of present interest. The factor $F_{(c)}$ comes from the non-resonant background
possibly associating with either the direct $\gamma^*\rightarrow D\bar{D}$ transition
or the nearby $\psi(2S)$ or other charmonium states.
Because the background contribution would be different in various channels,
the line shapes of the $\psi^{'}$ would not be the same in other channels,
e.g. $\psi^{'} \to p\bar{p}$ \cite{BES-plb735} and $p\bar{p}\pi^0$ \cite{BES-prd90}.
This is obviously true for other hadron states. However, here we do not dig into
this issue and parameterize $F_{(c)}$ as $F_{(c)}(s)=1/(s-m^2_{bg}+im_{bg}\Gamma_{bg})$
for simplicity. It should be pointed out that the $F_{(c)+\psi^{'}}(s)$
could be parameterized in other format, e.g.
the coupled-channel models \cite{Achasov,Achasov2013,Caopsi2014},
however at the price of more complex.

The measured $R^{ex}_{uds(c)+\psi^{'}}$ values versus c.m. energies
are taken from BESIII's report \cite{BES-prl97-262001},
as shown in Fig. \ref{fig1} with statistical error bars.
We fit these $R^{ex}_{uds(c)+\psi^{'}}(s)$ values at each energy point
to the theoretical formula described above using the least squares fitting method.
The objective function of the least squares to be minimized in the fit is defined as
\begin{equation}
\chi^2=\sum\limits^{68}_{i=1}\left(\frac{R^{ex}_{uds(c)+\psi^{'}}(s_i)-
R^{th}_{uds(c)+\psi^{'}}(s_i)}{\Delta{R^{ex}_{uds(c)+\psi^{'}}(s_i)}}\right)^2 \quad,
\end{equation}
where $R^{ex}_{uds(c)+\psi^{'}}(s_i)$ is the measured value with the statistical error
$\Delta{R^{ex}_{uds(c)+\psi^{'}}(s_i)}$ at the c.m. energy $s_i$, and
$R^{th}_{uds(c)+\psi^{'}}(s_i)$ is the corresponding theoretical value calculated by
Eq. (\ref{eq:Rseperate2}).

In the considered narrow energy range, the $R_{uds}$ could be viewed as a constant,
independent of the energy. The $\Gamma_{nonD\bar{D}}$ in Eq. (\ref{eq:EDwidth}) tends to be in the range of
0 $\sim$ 5 MeV with large uncertainty in various fitting strategies and we do not include it into the following fits.
So we have seven free parameters ($R_{uds}$, $q$, $m_{\psi^{'}}$,
$g_{\psi^{'}D\bar{D}}$, $g_{\psi^{'}\gamma}$, $m_{bg}$ and $\Gamma_{bg}$) in total.
Because $g_{\psi^{'}\gamma}$ could be determined by the leptonic width
$\Gamma_{\psi^{'}e^{+}e^{-}}=4\pi\alpha^2g^2_{\psi^{'}\gamma}/(3m^3_{\psi^{'}})$,
we perform two separate fits to the data. One of them (Fit i) is to fix $g_{\psi^{'}\gamma} = 0.2523$
by the $\Gamma_{\psi^{'}e^{+}e^{-}} = 0.262$ KeV in Particle Data Group \cite{PDG},
and the other is to let it being a free parameter (Fit ii).
The curves in Fig. \ref{fig1} show these fits, where the solid line is Fit i
and the dashed line is Fit ii. The corresponding fitted parameters are shown
in Table \ref{table1}, in which the errors are only statistical ones.
The achieving $\chi^2/ndf=$ 1.38 and 1.23 respectively for Fit i and ii
are obviously smaller than BESIII' result $\chi^2/ndf=94/61=1.54$ \cite{BES-plb652}.
Particularly, Fit ii gives a dip around 3.82 GeV, which describes the data perfectly well.
It gives $R_{uds}=2.156\pm 0.022$, whose central value is in excellent agreement
with the prediction of pQCD \cite{rhad}  and can directly be used to evaluate
the strong coupling constant $\alpha(s)$ at the energy scale of around 3 GeV.
However, the errors of the $R_{uds}$ are on the same level of BESIII' result,
and it still has some room for the improvement of the fit quality.
These are probably due to the big uncertainties of the data, whose systematic errors
are around the same scale of the statistical errors and not included here yet \cite{BES-prl97-262001}.
As a reference, the $\chi^2/ndf$ could be close to 1.0 within above formula
in a similar fit to the data of $e^+e^- \rightarrow D\bar{D}$ reactions \cite{Caopsi2014}.

\begin{table}[htp]
\centering
\begin{tabular}{llll}
\hline
   & Fit i & Fit ii   \\
\hline
$R_{uds}$                        &  2.165 $\pm$ 0.024 &  2.156 $\pm$ 0.022 \\
q                                &  1.58 $\pm$ 0.31   & -0.19 $\pm$ 0.21  \\
$m_{\psi^{'}}$ (MeV)             & 3784.4 $\pm$ 2.7   &  3816.0 $\pm$ 13.9 \\
$g_{\psi^{'}D\bar{D}}$           &  14.0 $\pm$ 0.8    &  14.1 $\pm$ 3.4 \\
$g_{\psi^{'}\gamma}$ (GeV$^{2}$) &  0.2523 (fixed)    & 0.417 $\pm$ 0.048  \\
$m_{bg}$ (MeV)                   & 3753.6 $\pm$ 4.6   &  3767.4 $\pm$ 2.6 \\
$\Gamma_{bg}$  (MeV)             & 37.9 $\pm$ 3.2     &  41.9 $\pm$ 6.0 \\
$\chi^{2}/ndf$                   &  85.52/62 = 1.38   &  74.94/61 = 1.23 \\
\hline
\end{tabular}
\caption{
Fitted parameters and achieving $\chi^{2}/ndf$ in Fit i and Fit ii, see text for details.}
\label{table1}
\end{table}

The parameter $q$ has big error in Fit ii. As can be seen in Eq. (\ref{eq:FanoFF}),
its value largely rests on the position of dip in the line shape,
which is however has large uncertainty. So it could deduce that the uncertainty of $q$
in Fit ii comes from the coincidence of the fitted $m_{\psi^{'}}$ and the dip position.
In addition, the sign of $q$ varies in Fit i and ii, and this is caused by the fitted
$m_{\psi^{'}}$ in these two fits are lying on the opposite sides of the dip position.

It is found that the fitted $m_{\psi^{'}}$ both in Fit i and ii are larger
than the BW values in PDG \cite{PDG}, even considering their big uncertainties.
Our obtained values should be treated as bare mass of the $\psi^{'}$ as argued
in Ref. \cite{CaoFano2014} and depend on the way of dealing with the background
term $F_{bg}$. However, the corresponding dressed mass would be close to
the PDG values and more sophisticated models are involved to extract its value
\cite{Caopsi2014}. The width $\Gamma_{\psi^{'}}\sim29$ MeV at the nominal mass
$m_{\psi^{'}}=3.773$ GeV calculated with the obtained $g_{\psi^{'}D\bar{D}}$
is consistent with the BW values in PDG. The obtained $g_{\psi^{'}\gamma}$
in Fit ii is bigger than that of the PDG value and the corresponding leptonic width
$\Gamma_{\psi^{'}e^{+}e^{-}}$ is around 2.7 times bigger than that of the PDG.
This is possibly due to the rough treatment of the parameterization of the $F_{bg}$.
In Eq. (\ref{eq:FanoFF}), the value of $g_{\psi^{'}\gamma}$ is directly relevant
to the form of $F_{bg}$.

The physical interpretation of $m_{bg}$ and $\Gamma_{bg}$ would be difficult
in present framework, though they are expected to be associated
to the parameters of $\psi(3686)$ state
\cite{HBLi,YJZhang,Achasov,Achasov2013,GYChen,Limphirat}.
However, it is impossible to accurately determine the mass and width
of the $\psi(3686)$ by the present data. It should be stressed that
the extracted $R_{uds}$ is stable and reliable regardless of the uncertainties
of these parameters, as long as the line shape of $\psi^{'}$ state
is correctly reproduced.

\begin{figure}[htp]
\centering
\includegraphics[width=0.5\textwidth]{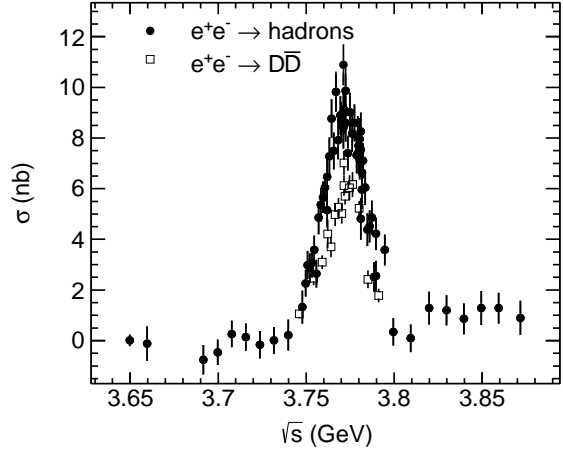}
\caption{
Cross sections of the $e^+e^- \rightarrow (c\bar{c})+\psi^{'} \rightarrow hadrons$
reaction extracted from measured R ratios at different c.m. energies compared to that of the
$e^+e^- \rightarrow D\bar{D}$ \cite{BES-plb668}.
}
\label{fig2}
\end{figure}

Using the $R_{uds}$ value extracted in Fit ii, we can obtain the cross sections
of $e^+e^- \rightarrow (c\bar{c})+\psi^{'} \rightarrow hadrons$ by Eq. (\ref{eq:Rseperate2}),
as shown in Fig. \ref{fig2}. The data of the $e^+e^- \rightarrow D\bar{D}$ cross section
from BESIII collaboration \cite{BES-prl97-121801,BES-plb668,BES-plb659} are plotted
in the same figure for comparison. The peak of $D\bar{D}$ production cross section
is obviously smaller and narrower than that of $hadrons$, which hints a non-zero $\Gamma_{nonD\bar{D}}$.
We calculate the cross section of $hadrons$ production to be $7.40\pm 0.69$ (stat.) nb at $\sqrt{s} =$ 3774.2 MeV,
which is bigger than the recent CLEO results $\sigma(e^+e^- \rightarrow D\bar{D})=6.489\pm 0.025 \pm 0.070$
at $\sqrt{s}=3774\pm 1$ MeV. However, they are still consistent with each other when both
the statistical and systematic uncertainties are taken into account.
Thus, the non-$D\bar{D}$ decay ratio of the $\psi^{'}$ is waiting for more precise measurements.

\section{Summary}

In short summary, we have performed a renewed analysis of the measured $R_{uds}$ value
from BESIII collaboration by treating the anomalous line shape of the $\psi(3770)$
resonance with a Fano-type formula. Our fitting results are better than
those in a simple Breit-Wigner resonance with energy dependent width, mainly because of
the improvement on the description of the dip structure at about 3.82 GeV.
The $R_{uds}$ value is determined to be $2.156\pm 0.022$ in the energy region between 3.650 and 3.872 GeV
from the data of BESIII Collaboration. The central value is consistent
with the pQCD calculation. We also reliably extract the cross sections of
the $e^+e^- \rightarrow hadrons$ without the continuum light hadron production,
which would be beneficial to our understanding of the properties of the $\psi(3770)$ state.

Our prescription of fitting the asymmetry line shape of states is
not only useful for pinning down the controversial decay ratios of $\psi(3770)$ state,
but also meaningful for determining the $R$ value in higher energy region
where is often overlapped with other resonances. The proposed framework is easily extended
to study other asymmetric line shapes of states and
could be served as a simple fitting strategy to the experimental data.

\section*{Acknowledgments}
One of the authors (X. Cao) would like to thank Prof. H. Lenske for useful discussion.
This work was supported by the National Basic Research Program (973 Program Grant No. 2014CB845406),
the National Natural Science Foundation of China
(Grant Nos. 11347146, 11405222, 11275235 and 11175220)
and Century Program of Chinese Academy of Sciences Y101020BR0.

\end{document}